\documentclass[12pt,reqno]{amsart}
\usepackage{amsmath,amsfonts,amssymb,amscd,amsthm,amsbsy,epsf}
\usepackage{cite}
\makeatletter
\@addtoreset{equation}{section}

\makeatother

\newtheorem{thm}{Theorem}[section]

\newtheorem{cor}[thm]{Corollary}
\newtheorem{prop}[thm]{Proposition}
\theoremstyle{remark}

\def\L{{\mathcal L}}
\def\F{{\mathcal F}}
\def\S{{\mathcal S}}
\def\G{{\mathcal G}}
\def\real{{\mathbb R}}

\begin{document}

\title[Boltzmann equations for mixture of Maxwell gases ]
{Boltzmann equations for mixtures of Maxwell gases: exact solutions and power like tails}

\author{A.V. Bobylev($^*$) and  I. M. Gamba($^{**}$)}

\address{\(^*\) Department of  Mathematics, Karlstad University, 
Karlstad, SE-651\,88 Sweden.}
\email{alexander.bobylev@kau.se}

\address{\(^{**}\) Department of Mathematics,
The University of Texas at Austin,
Austin, TX 78712-1082 U.S.A. }
\email{gamba@math.utexas.edu}

\begin{abstract} 
We consider the Boltzmann equations for mixtures of Maxwell gases.
It is shown that in certain limiting case the equations admit self-similar 
solutions that can be constructed in explicit form. More precisely, the solutions have 
 simple explicit integral representations. The most interesting solutions have 
finite energy and power like tails. This shows that power like tails can appear not just for
granular particles (Maxwell models are far from reality in this case), but also 
in the system of particles interacting in accordance with laws of classical mechanics. 
In addition, non-existence of positive self-similar solutions with finite moments of any order 
is proven for a wide class of Maxwell models. 

\end{abstract}

\maketitle
\begin{center}
{\it To Carlo Cercignani, on his 65th Birthday}
\end{center}

\bigskip

\section{Introduction}

In this paper we continue the study of self-similar solutions for various physical systems
described by the Boltzmann equations with Maxwell collision kernels  \cite{1,2,3,4}. 
It was understood long 
ago  \cite{5} that the key mathematical tool for such equations is the Fourier transform
 in the 
velocity space. A detailed analytical theory of the classical spatially homogeneous
Boltzmann equation for Maxwell molecules was mainly completed in the 80's (see 
 \cite{6} for a review). There were almost no new essential results in this field during the 
90's, except for some interesting pure mathematical publications, in particular \cite{7,8}. 

Quite unexpectedly,  Maxwell models again became the subject of many publications 
in the beginning of the 2000's. The starting point was  an idea to use such models 
for descriptions of inelastic (granular) gases. Inelastic Maxwell models were introduced
in 2000 \cite{1} 
(see also \cite{9} for the one dimensional case). It was clear that all the analytical 
techniques previously developed for classical Maxwell models can be used  in the inelastical 
case almost without changes. Many references to papers published in 2000s by physicists 
can be found in the book \cite{10}.

One interesting result (absent in the elastic case)  was the appearance of self-similar solutions 
with power like tails. It was conjectured in \cite{11} and later proved in \cite{3,4},
 that 
such solutions represent asymptotic states for a wide class of initial data. 
On the other
 hand, inelastic Maxwell models are just a rough approximation for the inelastic
 hard sphere model and they
give usually wrong answers to the question of large velocity asymptotics 
(a mathematically rigorous study of such asymptotics for a hard sphere model 
can be found in \cite{12}.)

Some new results in the theory  of classical (elastic) Boltzmann equation 
for Maxwell molecules were also recently published 
in \cite{2,3}: Self-similar solutions (two of which were found in explicit form)
 and the proof that such solutions represent a large time asymptotics for
 initial data with infinite energy, clarification of the old Krook-Wu 
conjecture \cite{13}, etc. It is clear that both elastic and inelastic
 Maxwell models must be studied from a unified point of view.

An interesting question arises in connection with power-like tails for high
velocities: Is it possible to observe a similar effect  (an appearance
of power-like tails  from initial data with exponential tails) in the system
of particles interacting according to laws of 
classical mechanics (i.e. without inelasticity assumption)?
This is the main question for the present paper. 

We shall see below that the answer to this question is probably affirmative.
 The key idea is to consider a mixture of classical Maxwell gases and to find a 
corresponding  limiting case for which such behavior can be in principle observed.

The paper is organized as follows. First we consider the Boltzmann equation for Maxwell 
mixtures and pass to the Fourier representation (Section 2). Then we study a binary 
mixture and show that corresponding equations formally admit a class of 
self-similar solutions (Section 3).
In order to simplify the problem we pass to the limit that corresponds to a one component gas
in the presence of the thermostat with fixed temperature $T$ (Section 4).
The general problem can be reduced to the case $T=0$ (cold thermostat). Then we consider 
the case   of the model cross section (pseudo-Maxwell molecules with isotropic
 scattering) and construct a family of exact self-similar solutions (Section 5). 
These solutions are studied in detail in Sections 6 and 7. 
The new  solutions have a lot in common with exact solutions from Ref. \cite{2}.
They, however, have finite energy and therefore are more interesting for 
applications.

We did not try to prove neither existence of such solutions in the more general case 
nor to show that they are large time asymptotic  states for a wide class of initial data.
 This is done, as a particular case,  in our  paper \cite{14} jointly with C. Cercignani.

Instead, we prove in Section 8  a general statement (applied, in particular, to inelastic 
Maxwell models and elastic models in the cold thermostat) 
concerning non-existence of positive self-similar solutions with finite moments of any order.

Thus it is sufficient in many cases to prove that  such positive solution does exist, 
then it always has just a finite number of even integer  moments for all values of 
parameters of  the equation.

\section{Maxwell mixtures}

We consider a spatially homogeneous mixture of $N\geq 2$ Maxwell gases. Each component
 of the mixture is characterized by the molecular mass $m_i$ and the distribution
 function  $f_i = f_i (v,t), \,i=1,\ldots,N$, where  $v \in\real^3$ and 
$t\in \real_+$ denote 
velocity and time respectively.
The distribution functions are normalized in such a way 
that  
\begin{equation*}\label{eq:2.1}
\int_{_{\real^3}} dv f_i(v,t)=\rho_i
\end{equation*}
where $\rho_i$ are the number density of the $i^{th}$ component of the mixture.
Note that the quantities $\rho_i, \,i=1,\ldots,N$ are preserved in time.

The  interaction between particles  is
described by the matrix of  Maxwell type differential cross-sections
\begin{equation*}\label{Maxwell-Molecule}
\sigma_{ij} (|u|,\theta) = \frac1{|u|} g_{ij} (\cos\theta) \ ,
\qquad i,j=1,\ldots, N\ .
\end{equation*}
where $|u|$ is the relative speed of colliding particles, $\theta\in[0,\pi]$
 is the scattering angle.

In case of  ``true'' Maxwell molecules, i.e. particles interacting with potentials
\begin{equation*}\label{eq19} 
U_{ij} = \frac{\alpha_{ij}}{r^4} \, ,\qquad 
\alpha_{ij} >0\ ,
\end{equation*}
where $r>0$ denotes a distance between interacting particles, the following formulas
are valid \cite{15}
\begin{equation*}\label{eq20} 
 g_{ij} (\cos \theta) 
= \left(\frac{\alpha_{ij}}{m_{ij}}\right)^{1/2} g(\cos\theta), \qquad \ \ 
m_{ij}= \frac{m_im_j}{m_i + m_j}\, ,\ \ \ i,j=1,\ldots,N\, .
\end{equation*}

We shall assume below the same kind of formulas for $g_{ij}(\cos\theta)$
with an arbitrary function  $g(\cos\theta)$ (pseudo-Maxwell particles).

The corresponding system of Boltzmann equations reads
\begin{equation}\label{Boltzmann}
\frac{\partial f_i}{\partial t} = \sum_{j=1}^N 
\int_{\real^3 \times S^{2}} dv_*\, d\omega\, g_{ij} 
\left(\frac{u\cdot 
\omega}{|u|}\right) 
\left[ f_i (v') f_j (v'_*) - f_i (v) f_j (v_*)\right]
\end{equation}
where the pair $(v',v'_*)$ are pre-collisional velocities
\begin{equation}\label{post-interacting}
\begin{split}
v' & = (m_i v + m_j v_* + m_j |u|\omega) (m_i + m_j)^{-1}\,  , \\
v'_* & = (m_i v + m_j v_* - m_i |u|\omega) (m_i + m_j)^{-1}\,  ;\qquad 
i,j=1,\ldots,N
\end{split}
\end{equation}
with respect to  the post-collisional velocities  $(v,v_*)$.

The Fourier transform of Eqs.~\eqref{Boltzmann}--\eqref{post-interacting}, for 
\begin{equation*}\label{Fourier}
\varphi_j (k,t) = \int_{_{\real^3}} dv\, f_j(v,t) e^{-ik\cdot v}\, ; \quad
k \in \real^3
\end{equation*}
leads to  equations \cite{6}
\begin{equation}\label{eq5}
\frac{\partial\varphi_i}{\partial t} = \sum_{j=1}^N \S (\varphi_i,\varphi_j)
\end{equation}
where 
\begin{equation}\label{eq6}
\S (\varphi_i,\varphi_j) 
= \int_{_{S^{2}}}\,d\omega\, g_{ij} \left(\frac{k\cdot\omega}{|k|}\right) 
\left[ \varphi_i (k_{ij}^+,t) \varphi_j (k_{ij}^-,t) 
- \varphi_i (k,t)\varphi_j (0,t)\right]
\end{equation}
where
\begin{equation*}\label{eq7}
k_{ij}^+ = \frac{m_i k + m_j|k|\omega}{m_i + m_j} \ ,\qquad
k_{ij}^- = \frac{m_j}{m_i + m_j}  (k-|k|\omega)
\end{equation*}
so, 
\begin{equation*}\label{eq8}
k = k_{ij}^+ + k_{ij}^-\ ; \ i,j=1,\ldots,N \, .
\end{equation*}

We note that 
\begin{equation*}\label{eq9}
(k_{ij}^-)^2 = 
 4\left( \frac{m_j}{m_i + m_j}\right)^2 |k|^2 s \, , \qquad
|k_{ij}^+ |^2 = 
 |k|^2 \left( 1- 4\frac{m_i m_j}{(m_i + m_j)^2} s\right) \; ,
\end{equation*}
where 
\begin{equation*}\label{eq10} 
s = \frac12 \left( 1-\frac{k\cdot \omega}{|k|}\right) \ , |s|\leq 1 .
\end{equation*}

Then, we consider isotropic solutions of Eqs. \eqref{eq5}-\eqref{eq6} 
\begin{equation*}\label{eq12} 
\varphi_i (k,t) = \tilde\varphi_i \left( \frac{|k|^2}{2m_i},t\right) \ ,
\end{equation*}
and obtain 
\begin{equation}\label{eq13} 
\frac{\partial}{\partial t} \tilde\varphi_i 
\left( \frac{|k|^2}{2m_i},t\right) 
= \sum_{j=1}^N \tilde\S (\tilde\varphi_i,\tilde\varphi_j)\ ,
\end{equation}
with 
\begin{equation*}\label{eq14} 
\begin{split}
\tilde\S (\tilde\varphi_i,\tilde\varphi_j) 
&= \int_{_0}^{^1} ds\, G_{ij} (s) 
\left[ \tilde\varphi_i \left( \frac{|k|^2}{2m_i}  (1-\beta_{ij} s),t\right) 
\right.\\
&\qquad  \left. 
\tilde\varphi_j \left( \frac{|k|^2}{2m_j}  \beta_{ij} \frac{m_j}{m_i} s,t
\right) - \tilde\varphi_i \left(\frac{|k|^2}{2m_i},t\right) 
\varphi (0,t)\right]
\end{split}
\end{equation*}
and
\begin{equation*}\label{eq15} 
\beta_{ij} = \frac{4m_i m_j}{(m_i + m_j)^2}\ ,\qquad 
G_{ij} = 4\pi g_{ij}{(1-2s)}
\end{equation*}
for $i,j=1,\ldots,N$.
We omit the tildes  and denote 
\begin{equation*}\label{eq16}
x = \frac{|k|^2}{2m_i} 
\end{equation*}
in each of Eqs. \eqref{eq13}. Then  the resulting set of equations  becomes 
\begin{equation*}\label{eq17} 
\frac{\partial}{\partial t} \varphi_i (x,t) 
= \sum_{j=1}^N \int_0^1  ds\, G_{ij} (s) 
\left[ \varphi_i (x(1-\beta_{ij}s)) \varphi_j (x\beta_{ij} s) 
- \varphi_i (x) \varphi_j (0)\right]
\end{equation*}
 where 
\begin{equation*}\label{eq18} 
0 \le \beta_{ij} = 
4 \frac{m_i m_j}{(m_i + m_j)^2} \le 1\ ,\qquad 
\beta_{ij} = \beta_{ji} \, , \ \ \beta_{ii}=1\, , \ \quad i,j=1,\ldots,N \, .
\end{equation*}

Therefore, the most general system of isotropic Fourier transformed 
 Boltzmann equations for 
Maxwell mixtures reads 
\begin{equation}\label{eq21} 
\frac{\partial\varphi_i}{\partial t} 
= \sum_{j=1}^N \gamma_{ij} \langle \varphi_i ((1-\beta_{ij} s)x)
\varphi_j (\beta_{ij} s\,x) - \varphi_i (x) \varphi_j (0)\rangle \, ,
\end{equation}
where, for any function $A(s), s\in [0,1]$,  
\begin{equation}\label{eq22}\begin{split}
\langle A(s)\rangle = \int_0^1 A(s) G(s)\,ds \, , \  \ \ 
 &\qquad  G(s) = 4\pi g(1-2s)\; ,\\
\text{ and} \ \ \ 
 \qquad \qquad \qquad \qquad \qquad \qquad &\qquad  \\
\gamma_{ij} = \sqrt{\frac{\alpha_{ij}}{m_{ij}}}\, ,\ \  
\beta_{ij} = \frac{4m_{ij}^2}{m_i m_j}\, ,   &\qquad 
 m_{ij} = \frac{m_i m_j}{m_i + m_j} \ ; i,j=1,\ldots,N \ .
\end{split}\end{equation}

\section{ Binary mixture}

We consider a special case $N=2$ in Eqs. \eqref{eq21}, \eqref{eq22}
and denote 
 \begin{equation*}\label{eq22.1}
m_1=M\, , \ \ m_2=m\, , \qquad \qquad  \varphi_1 =\rho_{_1}\varphi (x,t)\, ,  
\varphi_2 = \rho_{_2} \psi (x,t)\, , 
\end{equation*}
such that    
$\varphi (0,t) = \psi (0,t) = 1$.
Then, we obtain
\begin{equation}\label{eq23} 
\begin{split}
\varphi_t & = \rho_{_1} \left( \frac{2\alpha_{11}}{M}\right)^{1/2} 
\langle \varphi,\varphi\rangle 
+ \rho_{_2} \left(\frac{\alpha_{12}}{m_{12}}\right)^{1/2} 
\langle \varphi,\psi\rangle_{_\beta}\\
\psi_t & = \rho_{_2} \left( \frac{2\alpha_{22}}{m}\right)^{1/2} 
\langle \psi, \psi\rangle 
+ \rho_{_1} \left(\frac{\alpha_{12}}{m_{12}}\right)^{1/2} 
\langle \psi,\varphi\rangle_{_\beta}
\end{split}
\end{equation}
where
\begin{equation*}\label{eq24}\begin{split} 
&\beta = \frac{4mM}{(m+M)^2}\, ,\quad \ m_{12} = \frac{mM}{(m+M)}\, , \\
&\langle\varphi,\psi\rangle_{_\beta} 
= \int_{_0}^{^1} ds \,  G(s) \left\{ \varphi ((1-\beta s)x,t) \psi (\beta s x,t) 
- \varphi (x,t)\psi (0,t)\right\}
\end{split}
\end{equation*}
following the notation of \eqref{eq22}, and thus,
\begin{equation*}\label{eq25}
\langle\varphi,\varphi\rangle 
= \langle\varphi,\varphi\rangle_1 \, ,\qquad  
\langle\psi,\psi\rangle 
= \langle\psi,\psi\rangle_1 \, .
\end{equation*}
We recall the connection of  functions  $\varphi (x,t) $ and $ \psi (x,t) $
with corresponding solutions $f_{1,2}(|v|,t)$ of the Boltzmann equations 
\eqref{Boltzmann}
\begin{equation*}\label{eq26}\begin{split} 
\rho_{_1}\varphi(\frac{|k|^2}{2M},t) = \int_{\real{^3}} dv\, f_1(|v|,t) 
e^{-i\,k\cdot v} \, ,\qquad\qquad
\rho_{_2}\psi(\frac{|k|^2}{2m},t) = \int_{\real{^3}} dv\, f_2(|v|,t) 
e^{-i\,k\cdot v} \, .
\end{split}
\end{equation*}

The usual definition of kinetic temperatures is given by equalities
\begin{equation*}
T_i= \frac{m_i}{3\rho_{_i}} \int_{\real{^3}} dv\,|v|^2 f_i(|v|,t) \, , \quad i=1,2\, .
\end{equation*}
Then, one can easily verify that 
\begin{equation*}\label{eq27}
T_1(t)= -\varphi'(0,t) ,\qquad  
T_2(t)= -\psi'(0,t) \, .
\end{equation*}
where the primes denote derivatives on $x$. The equilibrium temperatures $T_{eq}$
of the binary mixture reads
\begin{equation*}
T_{eq} =\frac{\rho_{_1}\,T_1 +\rho_{_2}\,T_2 }{\rho_{_1} +\rho_{_2}} =\text{ const.}
\end{equation*}

The relaxation process in the binary mixture described by Eqs. \eqref{eq23}
leads to usual Maxwell asymptotics states
\begin{eqnarray*}
\varphi \rightarrow_{t\to\infty} \exp(-T_{eq}\,x) \, , \qquad \psi \to_{t\to\infty} \exp(-T_{eq}\,x) \, .
\end{eqnarray*}

By using Eqs. \eqref{eq23}, one can easily 
verify (at the formal level) that
\begin{equation*}\begin{split}
\frac{dT_1}{dt} = -\lambda \rho_{_2}(T_1-T_2)\, , 
&\qquad\frac{dT_2}{dt} = -\lambda \rho_{_1}(T_2-T_1)\, , \\
\lambda=\left(\frac{\alpha_{12}}{m_{12}}\right)^{1/2}\, \beta \langle s\rangle\, ,
&\qquad\langle s\rangle = \int_0^1 ds\, G(s) \, s \; .
\end{split}\end{equation*}
Therefore,
\begin{equation}\begin{split}\label{eq28}
T_1(t) = T_{eq} + \frac{\Delta}{\rho_{_1}}\,e^{-\Lambda \,t}\, , 
&\qquad T_2(t) = T_{eq} + \frac{\Delta}{\rho_{_2}}\,e^{-\Lambda \,t}\, , \\
\Delta = \frac{\rho_{_1}\, \rho_{_2}}{\rho_{_1} + \rho_{_2}} \left(T_1(0)-T_2(0) \right) \, ,
&\qquad\Lambda=\lambda\,(\rho_{_1} + \rho_{_2}) \, .
\end{split}\end{equation}

It is easy to see that Eqs. \eqref{eq23} formally admit the following class of 
self-similar solutions
\begin{eqnarray*}\label{eq29}
\varphi(x,t) =\Phi(x\, e^{-\Lambda \,t}) \, e^{-T_{eq} \,x} , 
\ \ \qquad \psi(x,t) =\Psi(x\, e^{-\Lambda \,t}) \, e^{-T_{eq} \,x} \, .
\end{eqnarray*}

It is, however, difficult to investigate such solutions (in particular,
 to prove that corresponding distribution functions are positive) in the
 most general case. Therefore we consider a simplified problem.

\section{Weakly coupled binary mixture }

If the masses $M$ and $m$ are fixed, then Eqs.~\eqref{eq23}
contain five positive parameters $\rho_{_i}\, , \alpha_{ij}\, , j=1,2$.
We shall consider below a special limiting case  of Eqs.~\eqref{eq23}
(weakly interacting gases) such that 
\begin{eqnarray}\label{eq30}
\alpha_{12}\to 0\, ,\quad \rho_{_2}\to \infty\, ,\qquad  \rho_{_2}\sqrt{\alpha_{12}} =
\mathrm{const} \, .
\end{eqnarray}

We assume that the other parameters $\rho_{_1}\, ,\alpha_{11}$ and  $\alpha_{22} $ remain constant and denote
\begin{equation}\label{eq31}\begin{split}
&\varphi(x,t) = \tilde\varphi(x,\tilde t)\, ,\psi(x,t) = \tilde\psi(x,\tilde t)\, ,
 \  \qquad \tilde t = \rho_{_1}
\left(\frac{2\alpha_{11}}{M}\right)^{1/2}\, t \: ,\\
&\qquad \qquad \theta = \frac{ \rho_{_2}}{ \rho_{_1}} 
\left(\frac{\alpha_{12} M}{2\alpha_{11}m_{12}}\right)^{1/2} = \mathrm{const}\, . 
\end{split}\end{equation}

Then we formally obtain (tildes are omitted below)
\begin{eqnarray}\label{eq32}
\varphi_t = \langle \varphi, \varphi \rangle + 
\theta  \langle \varphi, \psi \rangle_{_\beta}\, , \qquad  \langle \psi, \psi \rangle=0\, ,
\end{eqnarray}
where it is assumed that the functions $\varphi(x,t), \psi(x,t) $ and 
their time derivatives  remain finite in the limit \eqref{eq30}. 
The limiting temperatures (see Eqs.~\eqref{eq28}) are given by the identities
\begin{eqnarray*}\label{eq33}
T_2(t) = T_2(0) =T_{eq}=\mathrm{const.},  \qquad\mathrm{and}  
\qquad T_1(t) = T_2(0) +
\left( T_1(0) - T_2(0)\right) e^{-\theta\beta\langle s\rangle t} 
\end{eqnarray*}
in the notation \eqref{eq31} (tildes are omitted).

Thus 
\begin{eqnarray*}\label{eq34}
 \langle \psi, \psi \rangle = 0\,, \ \  \psi(0,t) = 1\, , \ \ 
 \psi'(0,t) = -T_2(0) \ \ \Longrightarrow \psi(x,t) =  e^{-T_2(0)\, x} \, ,
\end{eqnarray*}
and  we reduce Eqs.~\eqref{eq32} to the unique equation for $\varphi(x,t) $
\begin{eqnarray}\label{eq35}
\frac{\partial\varphi}{\partial t} = \langle \varphi, \varphi \rangle + 
\theta  \langle \varphi,  e^{-T_2(0)\, x} \rangle_{_\beta}\, .
\end{eqnarray}

The general case $T_2(0)>0$ can be reduced to the case $T_2(0)=0$
by substitution
\begin{eqnarray*}\label{eq36}
\varphi(x,t) =\tilde{\tilde{ \varphi}}(x,t) \exp(-T_2(0)\, x) \, .
\end{eqnarray*}

Eq.~\eqref{eq35} shows that, in the limiting case \eqref{eq30}, the second 
component of the mixture plays a role of a thermostat with the fixed temperature 
 $T_2(0)$, moreover, it is enough to consider the case  $T_2(0)=0$ 
(cold thermostat). Then Eq.~\eqref{eq35}, in explicit form, reads
\begin{equation}\begin{split}\label{eq37}
\frac{\partial\varphi}{\partial t}=&\int_{_0}^{^1} ds\, G(s) \left\{
 \varphi(sx)\,  \varphi[(1-s)x] + \theta  \varphi[(1-\beta\,s)x]
- \varphi(x) \left[  \varphi(0) +\theta \psi(0)\right] \right\} \, , \\
&\qquad  \varphi(0)=\psi(0)=1 \, ,
\end{split}\end{equation}
where the argument $t$ of $\varphi(x,t)$ is omitted.

We  consider below Eq.~\eqref{eq37} assuming that $ \varphi(|k|^2, 0)$
is a characteristic function (Fourier transform of a probability measure in $\real^3$).
Then  Eq.~\eqref{eq37} describes a homogeneous cooling process in the system
of particles that interact between themselves and with the cold thermostat.
Though all interactions are elastic (at the microlevel), such system has much 
in common with the gas of inelastic particles considered in \cite{3,4}. 
It was proved 
in these papers that the general inelastic Maxwell model has the self-similar
asymptotics in a  certain precise sense. We conjecture the same asymptotic
 property
for solutions of Eq.~\eqref{eq37} (see   \cite{14} for its proof).
 In this paper we construct some explicit 
examples of self-similar solutions and show that such solutions have
 power-like 
tails for large velocities.


\section{Exact solutions in the Fourier-Laplace representation}

We consider Eq.~\eqref{eq37} with $\beta=1$ (equal masses $m=M=1$) and 
$G(s)=1$ (isotropic scattering). Then  Eq.~\eqref{eq37} reads
\begin{equation}\begin{split}\label{eq38}
\frac{\partial\varphi}{\partial t}=\int_{_0}^{^1} ds\,  
  \varphi[(1-s)x] [\varphi(sx) +  \theta]  -  (1+\theta) \varphi(x) 
 \, ,\qquad  \varphi(0)=1 \, .
\end{split}\end{equation}

From the physical point of view, this is a model for a mixture of two weakly interacting 
gases, which consist of ``particles'' with identical masses. One can assume that the sorts
 of ``particles'' differs, say, by color.
 
Our goal is to describe a family of self-similar solutions of 
Eq.~\eqref{eq38} such that 
\begin{equation}\label{eq:b2}
\varphi (x,t) = \psi (xe^{-\mu t})\ ,\quad 
\psi (x) \cong 1 - ax^p\ ,\qquad x\to 0\, , 
\end{equation}
The parameters  $ \theta\geq 0$, $p>0$ and $\mu \in \real$ will be determined later. 
From now on $x$ denotes the self-similar variable. 
(The notation $ \psi (xe^{-\mu t})$ should not be confused with one for the 
function $\psi(x,t)$ from Sections 3 and 4.)

Substituting Eq.~\eqref{eq:b2} into Eq.~\eqref{eq38} 
we obtain
\begin{equation}\begin{split}\label{eq44}
\mu x \psi' (x)  - (1+\theta) \psi (x)  +\frac1x \psi * 
(\psi + \theta) =0\ ,\qquad
\psi_1 * \psi_2 = \int_0^x dy\, \psi_1 (y) \psi_2 (x-y)\ .
\end{split}\end{equation}

This equation can be simplified by the use of 
 the Laplace transform similarly to \cite{2} 
\begin{equation}\label{eq45}
w(z ) = \L (\psi) (z) 
= \int_0^\infty \psi (x) e^{-zx} \,dx\ ,\qquad
\text{Re } z >z_0\ .
\end{equation}
provided $|\psi (x)| < A \exp (z_0x)$, with some positive $A$ and $z_0$.

First, we recall  properties of the Laplace transform 
\begin{equation*}
\L (x\psi) = -w' (z)\ ,\quad
\L (x^2 \psi) = w'' (z)\ \text{ and }\ \L(\psi' ) = z\,w(z) - \psi (0) \, ,
\end{equation*}
and
\begin{equation*}
\L (x^2 \psi') = \frac{d^2}{dz}  (zw - \psi (0)) 
= (zw(z))'' \, .
\end{equation*}

Then we obtain the  following equation for $w(z)$:
\begin{equation*}\label{eq:2.2}
\mu (zw)'' + (1+\theta) w' + w\left( w + \frac{\theta}z\right) =0 \, .
\end{equation*}

Next, we denote by 
\begin{equation}\label{eq:b3}
u(z) = zw(z) = \int_0^\infty dx\, e^{-x} \psi \left(\frac{x}z\right)\ ,
\end{equation}
so that the above equation is transformed to 
\begin{equation}\label{eq:b4} 
\mu z^2 u'' + (1+\theta) z u' + u(u-1) =0\ .
\end{equation}

The  next step is to simplify this equation 
by standard substitutions. 
We denote 
\begin{equation}\label{eq:b5}
\tilde z = z^q\ ,\quad 
u(z) = \tilde u (\tilde z)\ ,
\end{equation}
and obtain the equation for $\tilde u (\tilde z)$ (tildes are omitted) 
\begin{equation*}
\mu q^2 z^2 u'' + q [1+\theta +\mu (q-1)] zu' + u (u-1) =0\ .
\end{equation*}
Then, setting
\begin{equation}\label{eq:b6}
u(z) = z^2 y (z) + B\ ,\qquad B = \text{const.},
\end{equation}
we obtain the following equation for $y(z)$: 
\begin{equation}\label{eq:b6.1}
\mu q^2 z^4 y'' + z^4 y^2 + \alpha z^3 y' + \beta z^2 y + B(B-1) =0\ ,
\end{equation}
where the parameters $\alpha$ and $\beta$ 
are given by the relations
\begin{equation*}
\begin{split}
\alpha & = q (5\mu q + 1+\theta -\mu)\ ,\\
\beta & = 2B-1 + 4\mu q^2 + 2q (1+\theta -\mu)\ .
\end{split}
\end{equation*}

Now, the parameters $q$ and $B$ can be chosen in such a way that 
$\alpha =\beta=0$.

Thus, we take
\begin{equation*}\begin{split}
q = -\frac{1+\theta -\mu}{5\mu}\ ,\qquad \qquad 
B = \frac{6\mu q^2 +1}2\ ,
\end{split}\end{equation*} 
and obtain that Eq.~\eqref{eq:b6.1} is reduced to
\begin{equation}\label{eq:b7}
\mu q^2 y'' + y^2 + \frac{B(B-1)}{z^4} =0\ .
\end{equation}

This is the simplest standard form, for which Eq.~\eqref{eq:b4} can be 
reduced in the general case. 
Finally, equation \eqref{eq:b7} is of the Painlev\'e type if and only if 
$B=0$ or $B=1$. 
Otherwise, it has moving logarithmic 
singularities and does not have   any 
``simple'' nontrivial analytic solutions (this is just a repetition of  
arguments on a similar equation considered in \cite{2}.) 

Therefore we consider the two special cases. In both cases
\begin{equation}\label{eq:b7.1}
 \theta =  \mu -1-5\mu q \, .
\end{equation}
\begin{itemize}
\item[{\bf Case 1:}] $B=0$, which implies 
\begin{equation}\label{eq:b7.2} 6\mu q^2 =-1,\qquad  \text{ so that } 
\qquad y'' = 6y^2\, .
\end{equation}
\item[{\bf Case 2:}]  $B=1$, which implies 
\begin{equation}\label{eq:b7.3} 6\mu q^2 =1,\qquad \text{ so that } 
\qquad y'' = - 6y^2 \, .
\end{equation}

\end{itemize}

The general solution of the  non-linear ODE $y'' = by^2$ 
is expressed in terms of 
Weierstrass elliptic function.
Similarly to \cite{2}, 
we can show that just the simplest exact  solutions, namely
\begin{equation*}\text{\bf Case
1:}  \quad y(z) = (c_1 +z)^{-2}\ \qquad  \text{and}
\qquad \text{\bf Case2:}\quad y(z) = -(c_2 +z)^{-2}\ ,
\end{equation*}
with  constant $c_{1}$ and $c_2 $, are solutions to Eqs.~\eqref{eq:b7.2} and 
 \eqref{eq:b7.3} respectively,
that lead to solutions of Eq.~\eqref{eq:b4} satisfying appropriate boundary 
conditions at infinity.

Coming back to the original notation
(see the transformations \eqref{eq:b5}, \eqref{eq:b6}), 
we obtain, therefore, two different exact solutions of 
Eq.~\eqref{eq:b4} 
\begin{equation}\label{eq:b7.4}
\begin{split}
&\text{{\bf Case 1:}}  \quad \ \ u(z) = (1+c_1 z^{-q})^{-2} \ \text{ for} 
\ \ 6\mu q^2 = -1\, ; \\ 
\ & \\
&\text{{\bf Case 2:}}  \quad \ \  u(z) =1 - (1+c_2 z^{-q})^{-2}  
\ \text{ for }\   6\mu q^2 =1 \, .
\end{split}\end{equation}
We have that, in both cases, the coupling constant $\theta$   must satisfy
\eqref{eq:b7.1}.

The constants $c_1$ and $c_2$ are determined by the boundary conditions as
 follows. The given asymptotics in
\eqref{eq:b2} for $\psi (x)$, $x\to0$, leads to 
the asymptotics for $u(z)$, as defined in \eqref{eq:b3},  at infinity
\begin{equation*}
u(z) \simeq 1  + \frac{b}{z^p}\ ,\qquad z\to\infty\ ,
\end{equation*}
where $b$ is a non-zero constant.
Recalling  that $|\psi (x)| \le 1$ for positive solutions of the 
Boltzmann equation, we can assume,  without loss of generality, that 
\begin{equation}\label{eq:b8}
u (z) \simeq 1 - \frac1{z^p}\ ,\qquad z\to\infty\ .
\end{equation} 

\bigskip

Then  we obtain, for the  above two cases in  \eqref{eq:b7.4},  the following 
formulas satisfying 
the boundary condition  \eqref{eq:b8} at  infinity: 
\begin{equation} \label{eq:b9}
\begin{split}
\text{{\bf Case 1:}}\  &q = p\, ;\quad  u(z) =  
\left( 1+\frac12 z^{-p}\right)^{-2}\, ,
     \quad   \mu = -\frac1{6p^2}\, ;\quad \theta = \frac{(3p-1)(1-2p)}{2p^2}\, . \\
&\ \\
\text{{\bf Case 2:}}\  &q  = -\frac{p}2\, ,\quad u (z) =1 - (1+z^{p/2})^{-2}\, , 
\quad
\mu  = \frac2{3p^2}\, ,\quad \theta = \frac{(3p+1)(2-p)}{3p^2}\, .
\end{split}
\end{equation}

The result can be formulated in the following way.

\begin{prop}\label{prop:b1}
Eq. \eqref{eq38} has exact self-similar solutions \eqref{eq:b2} 
satisfying the condition 
\begin{equation*}
\psi (x) \simeq 1 - \frac{x^p}{\Gamma (p+1)}\ ,\qquad x\to 0\ ,\quad p>0\ ,
\end{equation*}
for the following values of the parameters $\theta (p)$ and $\mu (p)$:
\begin{equation}\label{eq:b11}
\begin{split}
\text{\bf 1:} \quad  \mu (p) &= - \frac1{6p^2}\, ,\quad 
\theta (p) = \frac{(3p-1)(1-2p)}{6p^2}\ ;\\
\text{\bf 2:}\quad  \mu (p) &= \frac2{3p^2}\, ,\qquad
\theta (p) = \frac{(3p+1)(2-p)}{3p^2}\ .
\end{split}
\end{equation}
The solutions of Eq.~\eqref{eq38} are given by equalities
\begin{equation}\label{eq:b12} 
\psi_i (x) = \L^{-1} \left[ \frac{u_i(z)}{z}\right]\, ,\qquad 
i = 1,2\ ,
\end{equation}
with $u_{1,2}(z)$ from Eqs.~\eqref{eq:b9}, for {cases~1} and { 2}  
respectively.
\end{prop}

The solutions have a physical meaning if, both, 
$\theta \ge 0$ and  $\psi (\frac{|k|^2}2)$ 
is the Fourier transform of a positive function (measure).
 
The first condition leads to inequalities $\frac13 \le p\le\frac12$ in the 
{ case~1}, and  to $0\le p \le 2$ in the { case~2}. 
The second condition will be discussed in the next section.

\section{Distribution functions}

First we evaluate the inverse Laplace transforms \eqref{eq:b12}. 
In  the case 1 we obtain 
\begin{equation*} 
\psi_1 (x) = \L^{-1} \left[ \frac1z \left( 1+\frac{z^{-p}}2\right)^{-2}\right]
\ ,\qquad \frac13 \le p\le \frac12\ .
\end{equation*} 
The general formula from \cite{17} leads to 
\begin{equation}\label{eq:b13} 
\begin{split}
\psi_1 (x) & = \Phi (2^{-1/p} x)\ ,\quad 
\Phi (x) = \L^{-1} \left[\frac1z  (1+z^{-p})^{-2}\right] \\
&= 2\frac{\sin p\pi}{p \pi} \int_0^\infty ds \, e^{-xs^{-1/p}}\, 
\frac{ (1+s\cos p\pi)}{(1+s^2 + 2s \cos p\pi)^{2}} \  .
\end{split}
\end{equation}
This case corresponds to 
\begin{equation}\label{eq:b14}
\theta = \frac{(3p-1)(1-2p)}{6p^2}
\end{equation}
in Eq.~\eqref{eq38}. 
We note that $\theta =0$ for $p =\frac13$, $\frac12$. 
Eq.~\eqref{eq38} in such cases is the Fourier transformed Boltzmann 
equation  for one-component gas. 
The exact solutions \eqref{eq:b13} with $p =\frac12$, $\frac13$ were 
already obtained in \cite{2}. 
Eq.~\eqref{eq:b13} therefore yields a generalization of these 
solutions to the case  of binary mixture (with equal masses) provided 
the parameter $\theta$ is given (for given $p \in [1/3,\,1/2]$) in 
Eq.~\eqref{eq:b14}. 
The corresponding solutions of the Boltzmann equations are positive 
solutions with infinite energy. 
All their properties can be studied in the same way as in \cite{2}. 

The case 2 is more interesting since it includes also solutions with 
finite energy. 
The inverse Laplace transform 
\begin{equation*}
\psi_2 (x) = \L^{-1} \left\{ \frac1z [ 1-  \frac1{(1+z^{p/2})^2} ]\right\}
\end{equation*}
can be evaluated in the following way. 
We denote
\begin{equation*}\label{eq:b15} 
\Psi (x) = \L^{-1} \left[ \frac1{(1+z^{p/2})^2}\right]\ ,
\end{equation*}
then 
\begin{equation}\label{eq:b16}
\psi_2 (x) = 1-\int_0^x dy\, \Psi (y) 
= \int_x^\infty dy\, \Psi(y)\ ,\qquad 
\int_0^\infty dy\, \Psi (y)=1\ .
\end{equation}
The function $\Psi(x)$ can be expressed through the integral
\begin{equation*}
\Psi (x)  = \frac1{2\pi i} \int_{_C} \frac{dz\,e^{xz}}{(1+z^{p/2})^2}\, , 
\qquad 0<p\le 2\ 
\end{equation*}
where the contour $C$ lies  around the negative half of the real axis 
(see \cite{17} for details).

Then we obtain 
\begin{equation*}
\Psi (x) = \frac1{\pi} \int_0^\infty dr\, e^{-rx} A(r)\ ,
\end{equation*}
where 
\begin{equation*}
A(x) = \text{Im} \left[ 1+ (re^{-i\pi})^{p/2}\right]^{-2} 
= \frac{2r^{p/2} (1+ r^{p/2} \cos \frac{p\pi}2) \sin \frac{p\pi}2}
{(1+ r^p + 2r^{p/2} \cos \frac{p\pi}2)^2}\ .
\end{equation*}

Coming back to the function $\psi_2 (x)$ \eqref{eq:b16}, we obtain 
\begin{equation}\label{eq:b17}
\psi_2 (x) = \frac1{\pi} \int_0^\infty \frac{dr\, e^{-rx} A(r)}{r}\ .
\end{equation}

The final result 
\begin{equation*}\label{eq:b18} 
\psi_2 (x) = \frac{4\sin \frac{p\pi}2}{p\pi} 
\int_0^\infty 
\frac{ds(1+s\cos \frac{p\pi}2)}{(1+s^2 +2s\cos \frac{p\pi}2)^2}
\ e^{-xs^{2/p}}\ ,\qquad 
0<p\le 2\ ,
\end{equation*}
is obtained by substitution $s=r^{p/2}$ in the integral \eqref{eq:b17}. 

We remind to the reader that the corresponding distribution function 
$f(|v|,t)$, that solves the Boltzmann equation, reads 
\begin{equation}\label{eq:b19}
f(|v|,t) = e^{3\mu t/2} F(|v|e^{\mu t/2})\ ,\qquad 
\mu = \frac2{3p^2}\ ,
\end{equation}
where 
\begin{equation*}
\F[F] = \int_{\real^3} dv\, F(|v|) e^{-i\, k\cdot v} 
= \psi_2 \left(\frac{|k|^2}2\right)\ .
\end{equation*}

Noting that 
\begin{equation*}
e^{-T\frac{|k|^2}2} 
= \F [M_T (|v|)]\ ,\qquad 
M_T (|v|) = \frac{e^{-\frac{|v|^2}{2T}}}
{(2\pi T)^{3/2}}\ ,
\end{equation*}
we obtain the integral representation of $F(|v|)$:
\begin{equation}\label{eq:b20} 
F(|v|) = \frac{4\sin \frac{p\pi}2} {p\pi} \int_0^\infty 
ds\frac{ (1+s\cos \frac{p\pi}2)} {(1+s^2 + 2s\cos \frac{p\pi}2  )^2}\ 
M_{s^{2/p}} (|v|)\ .
\end{equation}
This function is obviously positive for $0<p\le 1$. 
We note that the solution \eqref{eq:b19}-\eqref{eq:b20} corresponds 
to the value 
\begin{equation*}\label{eq:b21} 
\theta  = \frac{(3p+1)(2-p)} {3p^2} 
\end{equation*}
in Eq.~\eqref{eq38}.

\section{Solutions with finite energy}

We consider in more detail the most interesting (for applications) 
case $p=1$ in Eq.~\eqref{eq:b20}. 
Then 
\begin{equation}\label{eq:b22} 
\mu =\frac23\ ,\quad \theta = \frac43\ ,\qquad 
f(|v|,t) = e^t F(|v|e^{t/3})\ ,
\end{equation}
where 
\begin{equation}\label{eq:b23} 
F(|v|) = \frac4{\pi} \int_0^\infty 
ds \,\frac{ \exp (- |v|^2/2s^2)} {(2\pi s^2)^{3/2} (1+s^2)^2}\ .
\end{equation}

We denote 
\begin{equation}\label{eq:b24} 
y = \frac{|v|^2}2\ ,\qquad
F(|v|) = (2\pi^5)^{-1/2} \Phi (y)\ ,
\end{equation}
where
\begin{equation}\label{eq:b25} 
\Phi (y) = \int_0^\infty dr \,\frac{ r^2}{(1+r)^2} \ e^{-ry}
= \frac1{y^3} \int_0^\infty dr\,\frac{ r^2 e^{-r}} {(1+\frac{r}y)^2}\ .
\end{equation}
Asymptotic expansion of $\Phi (y)$ for large positive $y$ follows from 
integration of the formal series 
\begin{equation*}
\left( 1+\frac{r}y\right)^{-2} 
= \sum_{n=0}^\infty (-1)^n (n+1) \left( \frac{r}y\right)^n\ .
\end{equation*}
Thus we obtain from Eq.~\eqref{eq:b25}
\begin{equation}\label{eq:b26}
\Phi (y) \simeq \frac1{y^3} \sum_{n=0}^\infty (-1)^n 
\frac{(n+1)(n+2)!}{y^n}\ ,\qquad 
y\to \infty
\end{equation}

In order to describe a behavior of $\Phi (y)$ for small positive $y$ 
we transform the first integral in Eq.~\eqref{eq:b25} in the 
following way 
\begin{equation*} 
\Phi (y) = \int_1^\infty dr \, \frac{(r-1)^2}{r^2} e^{-(r-1)y}
= e^y [E_0 (y) - 2E_1 (y) + E_2 (y)]\ ,\quad 
E_m (y) = \int_1^\infty dr\,\frac{ e^{-ry}}{r^m}\ .
\end{equation*}
Noting that 
\begin{equation*}
E_0 (y) = \frac{e^{-y}}y\ ,\quad 
E_2 (y) = e^{-y} - yE_1 (y)\ ,
\end{equation*}
we obtain 
\begin{equation*}
\Phi (y) = 1+\frac1y - (2+y) e^y E_1 (y)\ ,
\end{equation*}
where 
\begin{equation*}
E_1 (y) = \int_y^\infty ds\, \frac{ e^{-s}}{s} 
= \int_1^\infty ds\,\frac{e^{-s}}{s} 
+ \int_y^1  ds\, \frac{ (e^{-s}-1)}{s} - \ln y\ .
\end{equation*}

We note that
\begin{equation*} 
\int_1^\infty ds\, \frac{ e^{-s}}{s} 
+ \int_0^1 ds\, \frac{ (e^{-s}-1)}{s} 
= \int_0^\infty ds\, e^{-s} \ln s = -\gamma\ ,
\end{equation*}
where $\gamma \simeq 0,577$ is the Euler constant. 
Therefore
\begin{equation*}
E_1 (y) = - (\gamma +\ln y) + \int_0^y ds\, \frac{(1-e^{-s})} {s}
\end{equation*}
and 
\begin{equation}\label{eq:b27} 
\Phi (y) = 1 +\frac1y + (2+y) 
\left[ e^y (\gamma +\ln y) - \int_0^y ds\, \frac{ (e^s-1)} {s} \right]\ .
\end{equation}
Thus,  the asymptotic equality from  \eqref{eq:b26} and the formula 
\eqref{eq:b27} describe the behavior of the distribution function 
$F(v)$ in \eqref{eq:b24}, for large and small values of $|v|$.
We obtain 
\begin{equation}\label{eq:b28} 
\begin{split} 
F(|v|) & = 2\left( \frac2{\pi}\right)^{5/2} 
\frac1{|v|^6} \left[ 1+ O\left(\frac1{|v|}\right)\right]\ ,\qquad 
|v|\to\infty\ ,\\
F(|v|) & = \frac{2^{1/2}}{\pi^{5/2}}\ \frac1{|v|^2} 
\left[ 1 + 2|v|^2\ln |v| + O (|v|^2)\right]\ ,\qquad |v|\to 0\ .
\end{split}
\end{equation}

\bigskip

All the exact solutions can be generalized to the case of the   thermostat
with finite temperature $T$ (see Eq.~\eqref{eq34} with $T_2(0)=T$). Then  
 Eq.~\eqref{eq38} is replaced by the following equation 
\begin{eqnarray*}
\frac{\partial\varphi}{\partial t}=\int_{_0}^{^1} ds\left\{
 \varphi(sx) \varphi[(1-s)x] - \varphi(x)\varphi(0) \right\}
+ \theta \int_{_0}^{^1} ds \left\{ \varphi[(1-s)x] e^{-Tsx}
-  \varphi(x) \right\}  ,
\quad  \varphi(0)=1 .
\end{eqnarray*}

This equation can be reduced to  Eq.~\eqref{eq38} by substitution  
\begin{eqnarray*}
 \varphi(x,t)= \hat\varphi(x,t) e^{-Tx} \, .
\end{eqnarray*} 
The corresponding self-similar solutions read
\begin{eqnarray*}
 \varphi(x,t)=\psi(x e^{-\mu t}) e^{-Tx} \, ,
\end{eqnarray*} 
where $\psi(x)$ satisfies  Eq.~\eqref{eq44}.

\section{Self-similar solutions and power like tails}

We consider in this section a more general class of equations 
for the function $ \varphi(x,t)$:
\begin{equation}\begin{split}\label{eq8.1}
&\frac{\partial\varphi}{\partial t}=\int_{_0}^{^1} ds\,G(s) \left\{
 \varphi(a(s)x)\,  \varphi[(b(s)x] - \varphi(x)\, \varphi(0) \right\}
+ \theta \int_{_0}^{^1} ds\, H(s)\left\{ \varphi[c(s)x] 
-  \varphi(x) \right\} \, , \\
&\varphi(0)=1 \, ,
\end{split}\end{equation}
with non-negative functions $G(s), H(s), a(s), b(s)$ and $c(s)$ with $s\in [0,1]$.
We also assume that  $G(s), H(s)$ are integrable on $[0,1]$, and 
\begin{equation*}\begin{split}\label{eq8.2}
a(s)\leq 1, \ \ b(s)\leq 1, \ \ c(s)\leq 1 \, ; \ \ 0\leq s\leq 1\, .
\end{split}\end{equation*}

The function  $ \varphi(x,t)$ is understood as the Fourier transform 

\begin{equation}\begin{split}\label{eq8.3}
\varphi_ (|k|^2,t) = \int_{_{\real^d}} dv\, f(|v|,t) e^{-ik\cdot v}\, ; \quad
k \in \real^d ,\ \  d=1,2,\ldots \, ,
\end{split}\end{equation}
of a  time dependent probability density $f(|v|,t) \in \real^d$.

Eq.~\eqref{eq8.1} allows to consider from a unified point of view two different
 kind of Maxwell models:

\begin{itemize}
\item[{\bf I}] Inelastic Maxwell models \cite{1,3},
 where
\begin{equation}\begin{split}\label{eq8.5}
H(s)=0 ,\ \ a(s)= sz^2 , \ \ b(s)= 1-sz(2-z), \ \ z=\frac{1+e}2, 
0\leq e\leq 1\, .
\end{split}\end{equation}

\item[{\bf II}] Maxwell mixtures described by Eq.~\eqref{eq37}, where
\begin{equation}\begin{split}\label{eq8.6}
H(s)=\theta G(s) ,\ \ a(s)= s , \ \ b(s)= 1-s, \ \ c(s) = 1-\beta s\, .
\end{split}\end{equation}
\end{itemize}

The condition of integrability of $G(s)$ and $H(s)$ can be easily weakened.
We do not do it here in order to simplify proofs. Our main goal in this section 
is to prove, roughly speaking, that self-similar solutions (distribution functions 
 $f(|v|,t) $ from Eq.~\eqref{eq8.3}) have power-like tails.
 More precisely, we are going to show that such distribution functions 
can not have finite moments of any order.

Eq.~\eqref{eq8.1} admits (formally) a class of self-similar solutions 
\begin{eqnarray}\label{eq8.4}
 \varphi(x,t)=\psi(x e^{-\mu t})\, , \qquad  \mu>0 \, ,
\end{eqnarray}  

where $\varphi(x,t) $ satisfies the equation 
\begin{equation}\begin{split}\label{eq8.7}
&-\mu x \psi'=\int_{_0}^{^1} ds G(s) \left[
 \psi(a(s)x)  \psi((b(s)x) - \psi(x) \right]+ \int_{_0}^{^1} ds  H(s)\left[ \psi(c(s)x) 
-  \psi(x) \right] \, , \\
&  \psi(0)=1\, ,\qquad \qquad \ \mu > 0 \, .
\end{split}\end{equation}

The corresponding function  $f(|v|,t) \in \real^d$ (see  Eq.~\eqref{eq8.3})
reads
\begin{eqnarray}\label{eq8.8}
 f(|v|,t)= e^{d\,\frac{\mu t}2} F(|v| e^{\frac{\mu t}2}) \, ,\qquad 
\psi(|k|^2) = \int_{_{\real^d}} dv\, F(|v|) e^{-ik\cdot v}\,  .
\end{eqnarray}  

Our goal is to prove the following general fact: if such a function $F(|v|)\geq 0$
(generalized density of a probability measure in $\real^d$)  
does exists, then it can not have finite moments of all orders. We assume the
 opposite and represent the integral in Eq.~\eqref{eq8.1}
as a formal series
\begin{equation*}\begin{split}
\psi(|k|^2) = \sum^\infty_{n=0} \frac{(-1)^n}{(2n)!} \alpha_n(d)\,  m_n\, |k|^{2n}  \ ,
\end{split}\end{equation*}
where
\begin{equation*}\begin{split}
&m_n=  \int_{_{\real^d}} dv\, F(|v|) |v|^{2n}\, ,\ \  \ n=0,1,\ldots \, ; \\
&\alpha_n(1)=1 \, , \quad  \alpha_n(d)= \frac1{|S^{d-1}|} 
\int_{_{S^{d-1}}} d\omega \, (\omega'\cdot\omega)^{2n}\, , \ \ \ \text{if}\ \  d \geq 2 \, .
\end{split}\end{equation*}
where $|S^{d-1}|$ is the usual  measure of the unit sphere $S^{d-1}$ in $\real^d$,
$\omega'\in {S^{d-1}}$ is an arbitrary unit vector.

Hence we obtain
\begin{equation}\begin{split}\label{eq8.9}
&\psi(|k|^2) = \sum^\infty_{n=0} {(-1)^n} \psi_n\, \frac{ |k|^{2n}}{n!}  \, ,\\
& \psi_0=1\, ,\ \ \psi_n= \frac{n!}{(2n!)}\alpha_n(d)\, m_n, \ \ \ n=1,2,\ldots \, .
\end{split}\end{equation}

The convergence of the Taylor series~\eqref{eq8.9} is irrelevant for our goals.
 The only important  point is that 
$\psi(x)$ is infinitely differentiable for all $0\leq x< \infty$ (see any textbook
in probability theory, for example \cite{16}) and 
\begin{equation}\begin{split}\label{eq8.10}
\psi^{(n)}(0) = \psi_n >0 \, ,  \ \ \ n=0,1,\ldots \, .
\end{split}\end{equation}

On the other hand, the equations for $\psi_n$ can be easily obtained by substitution of 
the series~\eqref{eq8.9} into Eq.~\eqref{eq8.7}.
Then 
\begin{equation*}\begin{split}\label{eq8.11}
 \psi_0=1\, ,\ \ \psi_1 \, [\mu-\lambda(1)] =0 \, ,
\end{split}\end{equation*}
\begin{equation*}\begin{split}\label{eq8.12}
 \psi_n \, [\mu \, n-\lambda(n)] + \sum_{k=1}^{n-1} \G(k,n-k) \psi_k\psi_{n-k} =0 \, ,
 \ n=2,3,\ldots \ ,
\end{split}\end{equation*}
where 
\begin{equation*}\begin{split}\label{eq8.13}
& \lambda(n) = \int_{_0}^{^1} ds\,G(s) [1- a^n(s)- b^n(s)]
+ \int_{_0}^{^1} ds\, H(s)[1- c^n(s)] \, ,\\
&\G(k,l)=  {{k+l}\choose{k}} \int_{_0}^{^1} ds\, G(s) \,  a^k(s) b^l(s) \, , 
 \qquad k,l=1,2,\ldots \ .
\end{split}\end{equation*}
Now we can use conditions ~\eqref{eq8.10}. First we obtain $\mu=\lambda(1)$ and recall that $\mu>0$ by assumption ~\eqref{eq8.4}. Then we note that 
$ \G(k,l)\geq 0 $ for all $ k,l=1,2,\ldots  $. Therefore 
\begin{equation*}\begin{split}
 \psi_n \, [-\mu \, n + \lambda(n)] \geq 0 \ \ \ \Longrightarrow
\quad \frac {\lambda(n)}{n}\geq \mu >0\, , \ \   n=2,3,\ldots \ .
\end{split}\end{equation*} 
On the other hand, 
\begin{equation*}\begin{split}
\lambda(n)  \leq \int_{_0}^{^1} ds [G(s) +H(s)] < \infty\, , 
\end{split}\end{equation*} 
and, therefore, we get a  contradiction.

Thus, the following statement has been proven.

\begin{prop}\label{prop:8}
Equation~\eqref{eq8.7}, where $ \mu >0, G(s), H(s) \in L_+[0,1],
0\leq a(s) \leq 1, 0\leq b(s)\leq 1,  0\leq c(s)\leq 1 \, ;  s\in [0,1]$, 
does not have infinitely differentiable at $x=0$ solutions satisfying 
conditions~\eqref{eq8.10}. 
\end{prop}

\begin{cor}\label{cor:8}
The corresponding probability density $F(|v|)$ from  ~\eqref{eq8.8}
cannot have finite moments of all orders.
\end{cor} 

We can now apply the result to the inelastic Maxwell models ~\eqref{eq8.5}
 and conclude that the similar statement proved in our first paper \cite{1}
on that subject (see \cite{1}, Section 5, Theorem~5.1 ) is valid not just for almost all, 
but for all values of the restitution coefficient $e\in[0,1]$. 
Consequently we can revise now a statement from [3] related to existence of
self-similar solutions with finite moments of any order for a countable set
of values of e from the interval set [0,1] 
(a possible logarithmic singularity was missing in
the sketch of  proof of  Theorem 7.2 in [3]).
In fact the solution constructed in \cite{3} has a finite number of moments for 
{\bf any} $ 0\leq e <1$, without exceptions.

On the other hand, the above Proposition~\ref{prop:8} can be applied to 
Maxwell mixtures ~\eqref{eq8.6}. It shows that any {\sl physical}
(i.e. with a positive  $F(|v|)$ in Eq.~\eqref{eq8.8}) solution of 
Eq.~\eqref{eq8.7} corresponds to the distribution function  $F(|v|)$ 
with a finite number of even integer  moments. 
The exact solution constructed in Section~5 can be considered 
as a typical example.

\section*{Acknowledgements}

The first author was supported by  grant 621-2003-5357 from the 
Swedish Research Council (NFR). The second 
author has been partially supported by NSF under grant 
DMS-0204568.
Support from the Institute for 
Computational Engineering and Sciences at the 
University of Texas at Austin is also gratefully 
acknowledged.

\end{document}